\documentclass[11pt]{article}
\usepackage{graphicx}
\usepackage{amssymb}

\newtheorem{lemma}{Lemma}
\newtheorem{proposition}{Proposition}

\def\proof{\noindent  {\underline {Proof}}. }
\def\square{ {\hfill \vrule height6pt width6pt depth1pt} \bigskip \medskip }

\begin{document}

\centerline{\Large On the Quantum Inverse Problem for the Closed Toda Chain.}

\vskip 1cm
\centerline{O. Babelon
\footnote{Member of CNRS.}}
\vskip .5cm
\centerline{Laboratoire de Physique Th\'eorique et Hautes Energies.
\footnote{L.P.T.H.E. Universit\'es Paris VI--Paris VII (UMR 7589),
 Bo\^{\i}te 126, Tour 16, $1^{er}$ \'etage,
 4 place Jussieu, F-75252 PARIS CEDEX 05}
}
\date{March  2003}

\vskip 5cm

\begin{abstract}
We reconstruct the canonical operators $p_i,q_i$ of the quantum closed Toda chain
in terms of Sklyanin's  separated  variables. 
\end{abstract}

\vfill
LPTHE-P03-07
\eject

\section{Introduction.}

The theory of classical integrable systems relies  on two main ingredients. One is group theory
which is used to manufacture Lax matrices as coadjoint orbits of loop groups, and the second one
is complex analysis on the spectral curve, $\Gamma$, which is used to effectively solve the models.

In fact, once $\Gamma$ is given to us, we only need  $g={\rm genus}(\Gamma)$ points on 
it to reconstruct everything. 
The divisor ${\cal D}$ of these $g$ points is called the dynamical divisor.
Its role is fundamental. 
For instance, under an integrable flow, the curve $\Gamma$ 
is fixed but the points of ${\cal D}$ move on it. 
The main  theorem of integrable systems states that the image of ${\cal D}$ by the Abel map,
which is a point of the Jacobian of $\Gamma$,
moves linearly under such flows. Another very important property, which has emerged gradually, is that
the coordinates of the points of ${\cal D}$  
form a set separated variables in the sense of the Hamilton-Jacobi theory \cite{HoKrPh02, BaBeTa03}.

In the quantum theory too, these separated variables known there as Sklyanin's variables,
play an important role \cite{Skly85}. 
It was observed recently that the quantum commuting Hamiltonians 
had a simple and general expression in terms of Sklyanin variables \cite{BaTa02}. Hence, it becomes natural
to try to set up a  quantization procedure of a classical integrable system 
by using these variables systematically.

In this note, as an example, we perform this quantization program in the case of the closed Toda chain.
We will be able to reconstruct the original quantum Toda variables in terms of the 
Sklyanin variables, see eq.(\ref{reconquant}).  So, even in this most studied system, the approach seems powerful enough
to provide new results.

But before attacking the specific example of the Toda chain, it is worth recalling a few general
facts about the classical theory of integrable systems.

Lax matrices built with the help of coadjoint orbits of loop groups lead to spectral curves of the 
very special form \cite{KrPh97, BaTa99, BaBeTa03}
\begin{equation}
\Gamma: R(\lambda, \mu) \equiv R_0(\lambda,\mu) + \sum_j R_j(\lambda,\mu) H_j = 0
\label{specurve}
\end{equation}
where the $H_j$ are the Poisson commuting Hamiltonians. The coefficients 
$R_j(\lambda,\mu)$ have a simple geometrical meaning. It turns out that varying
 the moduli $H_i$ at $\lambda$ constant one can show that \cite{KrPh97}
\begin{equation}
\delta \mu \,d\lambda = - \sum_j {R_j\over  \partial_\mu R } \delta H_j = {\rm holomorphic}
\label{deltamu}
\end{equation}
hence 
$$
R_j (\lambda,\mu) =  N_j(\lambda,\mu)
$$
where $N_j$ are the numerators of the holomorphic differentials on $\Gamma$:
$$
\omega_j = {N_j(\lambda,\mu) \over \partial_\mu R (\lambda,\mu) } d\lambda = \sigma_j(\lambda,\mu) d\lambda
$$
The great virtue of eq.(\ref{deltamu}) is that it implies that there are exactly $g$ independent Hamiltonians because the space of 
holomorphic differentials is of dimension $g$. This is a most welcome fact because the natural candidates 
for the angles variables are the $g$ angle on the (complex) Jacobian torus, and so we need also $g$ (complex) action variables.  
This counting argument still holds if we  generalize eq.(\ref{deltamu}) as follows
\begin{equation}
{\delta \mu \,d\lambda\over f(\lambda,\mu)}   = {\rm holomorphic},\quad {\rm then~~} {R_j(\lambda,\mu) \over \partial_\mu R (\lambda,\mu) } = f(\lambda,\mu) \sigma_j(\lambda,\mu)
\label{delatmugeneral}
\end{equation}
Note that if we consider $\tilde{R}(\lambda,\mu) = h(\lambda,\mu) R(\lambda,\mu)$ where $ h(\lambda,\mu)$ does not contain 
dynamical moduli, we have
$$
 {\tilde{R}_j\over  \partial_\mu \tilde{R} }  = {h R_j\over  h\partial_\mu R +R\partial_\mu h } 
$$
so that {\em on $\Gamma$} the factor $h$ disappears and the factor $f$ above has an intrinsic meaning.

The $g$ moduli $H_i$  in eq.(\ref{specurve}), are completely determined if we require that 
the curve passes through $g$ points $\gamma_{k} = (\lambda_k,\mu_k)$. Indeed we just have to solve
the linear system
\begin{equation}
 \sum_{j=1}^g R_j(\lambda_k,\mu_k) H_j + R_0(\lambda_k,\mu_k) = 0,\quad k=1,\cdots ,g
\label{baxterclassique}
\end{equation}
whose solution is 
\begin{equation}
 H = - B^{-1} V
\label{hamiltoniensclassiques}
\end{equation}
where
$$
H = \pmatrix{H_1 \cr \vdots \cr H_i \cr \vdots \cr H_g},\quad
B = \pmatrix{ 
 R_1(\lambda_1,\mu_1) & \cdots & R_g(\lambda_1,\mu_1) \cr
\vdots   &     &  \vdots  \cr
 R_1(\lambda_i,\mu_i) & \cdots & R_g(\lambda_i,\mu_i) \cr
\vdots   &     &  \vdots  \cr
 R_1(\lambda_g,\mu_g) & \cdots & R_g(\lambda_g,\mu_g) },\quad
V = \pmatrix{ R_0(\lambda_1,\mu_1) \cr \vdots \cr  R_0(\lambda_i,\mu_i)
\cr \vdots \cr  R_0(\lambda_g,\mu_g)  }
$$

On the $2g$ complex numbers $(\lambda_k,\mu_k)$, we can introduce a non degenerate Poisson structure
\begin{equation}
\{ \lambda_k,\lambda_{k'}\}=0, \quad \{ \lambda_k,\mu_{k'}\} = p(\lambda_k,\mu_k) \delta_{kk'},
\quad \{ \mu_k,\mu_{k'}\}=0
\label{defpoisson}
\end{equation}
We do not need to specify the function $p(\lambda,\mu)$ for the moment. By a
simple calculation, we prove the \cite{ER01, BaTa02}
\begin{proposition}
For any function $p(\lambda,\mu)$ in eq.(\ref{defpoisson}), the Hamiltonians 
defined by eq.(\ref{hamiltoniensclassiques}) are in involution
$$
\{H_i,H_j\} = 0
$$
\end{proposition}
The $H_i$  therefore define integrable flows on the $2g$ dimensional phase space
eq.(\ref{defpoisson}). 

There is an interesting relation between the functions $p(\lambda,\mu)$ entering the Poisson bracket,
eq.(\ref{defpoisson}),
and the function $f(\lambda,\mu)$ in eq.(\ref{delatmugeneral}). 
\begin{proposition}
If $f(\lambda,\mu) = p(\lambda,\mu)$,  the flows linearise on the Jacobian.
\end{proposition}
\proof
Define the angles as the images of the divisor $\gamma_i$ by the Abel map:
$$
 \theta_j = \sum_k \int^{\lambda_k}  \sigma_j(\lambda,\mu) d\lambda
$$
We will show that
$$
\{ H_i,\theta_j\} = \delta_{ij}\quad {\rm i.e.~~} \sum_k \partial_{t_i} \lambda_k \;\sigma_j(\lambda_k,\mu_k) = \delta_{ij}
$$
Indeed, we have
\begin{eqnarray*}
\partial_{t_i} \lambda_k = \{ H_i, \lambda_k \} &=& - \{ B_{il}^{-1} V_l , \lambda_k \} \\
&=& B_{ir}^{-1} \{ B_{rs} , \lambda_k \} B_{sl}^{-1} V_l - B_{il}^{-1} \{ V_l, \lambda_k\} \\
&=& - B_{ik}^{-1} \Big[ \{ B_{ks} , \lambda_k \} H_s + \{ V_k , \lambda_k \} \Big]
\label{interesting}
\end{eqnarray*}
where in the last line, we used the separated structure of the matrix $B$ and the vector $V$. Explicitly
\begin{eqnarray*}
\partial_{t_i} \lambda_k B_{kj} &=&  B_{ik}^{-1} B_{kj}
\Big[ \partial_\mu R_s(\lambda_k,\mu_k)  H_s  +  \partial_\mu R_0(\lambda_k,\mu_k)  \Big]  p(\lambda_k,\mu_k) \\
&=&   B_{ik}^{-1} B_{kj}  \partial_\mu R(\lambda_k,\mu_k)  p(\lambda_k,\mu_k) 
\end{eqnarray*}
It follows that
$$
\partial_{t_i} \lambda_k \;{ R_j(\lambda_k,\mu_k) \over  p(\lambda_k,\mu_k) \partial_\mu R(\lambda_k,\mu_k) }
=  B_{ik}^{-1} B_{kj}
$$
summing over $k$ gives
\begin{equation}
\sum_k \partial_{t_i} \lambda_k \;{f(\lambda_k,\mu_k) \over  p(\lambda_k,\mu_k)} \sigma_j(\lambda_k,\mu_k) = \delta_{ij}
\label{disturbing}
\end{equation}
which reduces to what we had to prove when $f(\lambda,\mu) =  p(\lambda,\mu)$.
\square

Hence eqs.(\ref{specurve}, \ref{delatmugeneral}) do provide us with integrable systems and
their solution. The main question in this approach is to go back from the separated variables $(\lambda_k,\mu_k)$
to the "original" variables, i.e., the ones entering the Lax matrix elements. 
A Lax matrix is a matrix $L(\lambda)$
depending rationally on $\lambda$ such that
$$
R(\lambda,\mu) = \det ( L(\lambda)-\mu) 
$$
A general strategy to construct it is as follows \cite{DuKrNo90, BaBeTa03}.
First, we  determine the size of the matrix $L(\lambda)$, by looking at the curve $\Gamma$ as a covering of the
$\lambda$-plane $(\lambda,\mu) \to \lambda$. The dimension of the matrix is just the number of sheets
of this covering.  Call $Q_i(\lambda =\infty, a_i)$,
$i=1\cdots N$ the point above $\lambda = \infty$. We can normalise $L(\lambda) = {\rm Diag}(a_1,a_2,\cdots a_N)$
 at $\lambda = \infty$. To each point $P(\lambda,\mu)$ of the curve $\Gamma$, not a branch point 
 of the covering $(\lambda,\mu) \to \lambda$, one can attach the one dimensional eigenspace of $L(\lambda)$ corresponding to the eigenvalue $\mu$. One can show that this extends to an analytic line bundle on $\Gamma$ with Chern class $g$. 
 The eigenvector $\Psi(P)$ at $P\in \Gamma$ can be presented as follows
 $$
 \Psi(P) = \pmatrix{1\cr \psi_2(P)\cr \vdots ,\cr \psi_N(P)},\quad [\psi_i] = Q_1-Q_i -{\cal D}
 $$
The function $\psi_i$ has a zero at $Q_1$, a pole at $Q_i$ and $g$ poles at a divisor ${\cal D}$ at
finite distance. By the Riemann--Roch theorem, this function exists and is unique for ${\cal D}$ generic.
We identify ${\cal D}$ to the divisor of the $g$ points $\gamma_k=(\lambda_k,\mu_k)$ above and
 construct the corresponding vector $\Psi(P)$.
Once this is done,
we consider the $N$ points $P_i$ above $\lambda$, and build the matrices
$$
\widehat{\Psi} = (\Psi(P_1),\cdots ,\Psi(P_N)), \quad \widehat{\mu} = {\rm Diag}(\mu(P_1),\cdots , \mu(P_N))
$$
The matrix $L(\lambda)$ is given by
$$
L(\lambda) = \widehat{\Psi}\; \widehat{\mu} \;\widehat{\Psi}^{-1}
$$
This is independent of the order of the points $P_i$, and is a rational function of $\lambda$.

This method gives a way, in principle, to reconstruct the Lax matrix starting only from the spectral curve
and the dynamical divisor on it, hence returning back to the original variables.
Of course, in concrete examples, some of the genericity assumptions made here may have to be modified, 
or shortcuts may be available,  but the 
general ideas remain the same.

\section{The Classical Toda Chain.}

The closed Toda chain is defined by the Hamiltonian \cite{FaTa86}

\begin{equation}
H =  \sum_{i=1}^{n+1}{1\over 2} p_i^2 + e^{q_{i+1}-q_i}
\label{hamtoda}
\end{equation}
where we assume that $q_{n+2} \equiv q_1$, and Poisson bracket
$$
\{q_i,q_j\}=0,\quad \{p_i,q_j\} = \delta_{ij},\quad \{p_i,p_j\}=0
$$

This is an integrable system. We associate to it a Lax matrix as follows.
Consider the $2\times 2$ matrices
$$
T_j(\lambda) = \pmatrix{\lambda + p_j  & -e^{q_j} \cr e^{-q_j} &0}
$$
and construct
\begin{equation}
T(\lambda) = T_{1}(\lambda)\cdots T_2(\lambda)T_{n+1}(\lambda)
\label{nparticles}
\end{equation}
We can write
\begin{equation}
T(\lambda) = \pmatrix{A(\lambda) & B(\lambda) \cr
                                  C(\lambda) & D(\lambda) }; \quad A(\lambda) D(\lambda) - B(\lambda) C(\lambda) =1
\label{TToda}
\end{equation}
$A(\lambda)$ is a polynomial of degree $n+1$, $D(\lambda)$ is  of degree $n-1$, and
$B(\lambda)$, $C(\lambda)$ are of degree $n$.
The spectral curve is defined as usual
\begin{equation}
\det (T(\lambda) - \mu) = 0 \quad  \equiv \quad \mu + \mu^{-1} - t(\lambda) = 0
\label{spectralcurve}
\end{equation}
where 
$$
t(\lambda)=A(\lambda) + D(\lambda) = \lambda^{n+1} + \sum_{j=0}^{n} \lambda^j H_j, \quad
H_{n} = P,\quad H_{n-1} = {1\over 2} P^2 - H
$$
where $P = \sum_i p_i$, and $H$ is given by eq.(\ref{hamtoda}). The $n+1$ quantities $H_j$ are conserved.
The curve eq.(\ref{spectralcurve}) is hyperelliptic. It can be written as
\begin{eqnarray}
s^2 =  t^2(\lambda) - 4, \quad {\rm with}\quad  s = 2\mu -t(\lambda)= \mu - \mu^{-1}
 \label{specbis} 
\end{eqnarray} 
The polynomial $t^2(\lambda)$
being of degree $2(n+1)$, the genus of the curve is $g = n$. The number
of dynamical moduli is $g=n$ in the center of mass frame $P  =0$.
In the following we therefore always consider  the 
system reduced by the translational symmetry.
We have
\begin{equation}
{\delta \mu \over \mu}\;d\lambda = { \delta t(\lambda) \over \mu - \mu^{-1}}  d\lambda = 
 {\delta t(\lambda) \over  s} d\lambda =\; {\rm holomorphic}
 \label{holo1}
\end{equation}
Asking that the curve eq.(\ref{spectralcurve}) passes through the $n$ points $(\lambda_i,\mu_i)$,
we get $n$ equations
$$
  t(\lambda_i) = \mu_i + \mu_i^{-1} 
$$
Their solution for the $n$ Hamiltonians $H_i$ may be cast conveniently in the from of
Lagrange interpolation formula:
\begin{equation}
t(\lambda) = t^{(0)}(\lambda) + t^{(1)}(\lambda)
\label{lagrange}
\end{equation}
where
\begin{equation}
t^{(0)}(\lambda) =  (\lambda + \sum_i \lambda_i) \prod_{i=1}^n (\lambda - \lambda_i), \quad
t^{(1)}(\lambda) =   \sum_i  \prod_{j\neq i} {\lambda-\lambda_j\over \lambda_i -\lambda_j}
 \;(\mu_i + \mu_i^{-1})
 \label{lagrangebis}
\end{equation}
The polynomial $t^{(0)}(\lambda)$ is of degree $n+1$, vanishes for $\lambda = \lambda_i$ and has no  $\lambda^n$ term.

We define the Poisson bracket  of the separated variables as (in agreement with eqs.(\ref{defpoisson},\ref{holo1})):
$$
\{\lambda_k,\lambda_{k'}\} =0,\quad \{\mu_k,\lambda_{k'}\}= \mu_k \delta_{kk'} , \quad  \{\mu_k,\mu_{k'}\}=0
$$
By the general result of \cite{ER01, BaTa02} the Hamiltonians $H_i$ obtained as coefficients of the polynomial $t(\lambda)$ 
in eq.(\ref{lagrangebis}) are in involution. Notice that the above Poisson bracket is the one matching 
the condition (\ref{holo1}) and leads to flows linearizing on the Jacobian of the spectral curve eq.(\ref{spectralcurve}).

To proceed, we reconstruct the Lax matrix. The curve eq.(\ref{spectralcurve})
is a two sheeted cover of the $\lambda$-plane. For a $2\times 2$ matrix of the form eq.(\ref{TToda})
the eigenvector is simple
$$
(T(\lambda) - \mu) \Psi = 0, \quad \Psi = \pmatrix{1\cr \psi_2 }, \quad \psi_2 = -{A(\lambda) - \mu \over B(\lambda)}
$$
The poles of $\psi_2$ at finite distance are above the zeroes  $\lambda_i$ of $B(\lambda)=0$ which is a polynomial
of degree $n$. The two points above $\lambda_i$ are $\mu_i^+ = A(\lambda_i)$, $\mu_i^- = D(\lambda_i)$ so that 
$\psi_2$ has a pole only on the second point. The points of the dynamical divisor are therefore
$$
(\lambda_i, D(\lambda_i)), \quad B(\lambda_i) = 0
$$
Given the points of the dynamical divisor, we reconstruct $A(\lambda)$ and $B(\lambda)$:
\begin{eqnarray*}
B(\lambda) &=& b_0 \prod_{i=1}^n (\lambda - \lambda_i) \\
A(\lambda) &=& (\lambda + \sum_{i=1}^n \lambda_i)  \prod_{i=1}^n  (\lambda - \lambda_i) 
+ \sum_i \mu_i {\prod_{j\neq i}^n  (\lambda - \lambda_j) \over \prod_{j\neq i}^n  (\lambda_i - \lambda_j)}
\end{eqnarray*}
Knowing $A(\lambda)$ and $B(\lambda)$ we reconstruct $C(\lambda)$ and $D(\lambda)$ by the trace and 
determinant conditions. These formulae were the basis of Sklyanin's work \cite{Skly85}, and of Smirnov's work
\cite{Smi98} and  \cite{Smi00} (with a different Poisson structure).

 To reconstruct the original degrees of freedom of the 
Toda chain, however,  is equivalent to reconstructing  the $(n+1)\times (n+1)$ Lax matrix: 
$$
L(\mu) = \sum_i p_i E_{ii} + \sum_{i=1}^n e^{{1\over 2}(q_{i+1}-q_i)}(E_{i,i+1} + E_{i+1,i}) + e^{{1\over 2}(q_{1}-q_{n+1})}(\mu  E_{n+1,1} +
\mu^{-1} E_{1,n+1})
$$
where $(E_{ij})_{kl} = \delta_{ik}\delta_{jl}$.
This matrix is such that
$$
\det ( L(\mu) - \lambda ) = \mu + \mu^{-1} - A(\lambda) - D(\lambda)
$$
Since $L(\mu)$ is of size $(n+1)\times (n+1)$, we look at the spectral curve
eq.(\ref{spectralcurve}) as a $n+1$ sheeted cover of the $\mu$-plane. 
When $\lambda = \infty$, we have two points $P^+$ and $P^-$ corresponding to $\mu = \infty$ and
$\mu = 0$ respectively:
\begin{eqnarray*}
P^+ &:& \mu = \lambda^{n+1} (1 + O(\lambda^{-2}))\\
P^- &:& \mu = \lambda^{-n-1} (1 + O(\lambda^{-2}))
\end{eqnarray*}
According to e.g. \cite{DuKrNo90, BaBeTa03}, the eigenvectors of $L(\mu)$ are easy to construct. Set
$$
\Psi = \pmatrix{\psi_1 \cr \psi_2 \cr \vdots \cr \mu }
$$
where we have normalised the last component to be $\mu$. The meromorphic functions 
$\psi_i$ have poles at the dynamical divisor, moreover 
\begin{eqnarray*}
\psi_i &=& e^{{q_i - q_{n+1} \over 2}} \lambda^i ( 1 + O(\lambda^{-1} ), \quad {\rm ~~~~near~ } P^+\\
\psi_i &=& e^{-{q_i - q_{n+1}\over 2}} \lambda^{-i} ( 1 + O(\lambda^{-1} ), \quad {\rm ~near~ } P^-
\end{eqnarray*}
These properties determine the functions $\psi_i$ uniquely. Being meromorphic functions
on a hyperelliptic curve, we can write
$$
\psi_i = { Q^{(i)}(\lambda) + \mu R^{(i)}(\lambda) \over \prod_{j=1}^n (\lambda - \lambda_j) }
$$
where $Q^{(i)}$ and $R^{(i)}$ are polynomials. We want the poles to be at $(\lambda_j, \mu_j)$ only 
so that the numerator should vanish at the points  $(\lambda_j, \mu_j^{-1})$. This gives
$n$ conditions
\begin{equation}
Q^{(i)}(\lambda_j) + \mu_j^{-1} R^{(i)}(\lambda_j) = 0, \quad j = 1 \cdots n
\label{conditions}
\end{equation}
To have a pole of order $i$ at $P^+$ and a zero of order $i$ at $P^-$ we choose
$$
{\rm degree~} Q^{(i)} = n-i, \quad {\rm degree~} R^{(i)} = i-1
$$
These two polynomials depend altogether on $n+1$ coefficients. They are determined by imposing
the $n$ conditions eq.(\ref{conditions}) and requiring that the normalizations coefficients are
inverse to each other at $P^\pm$. We set
\begin{eqnarray*}
Q^{(i)}(\lambda) &=& Q^{(i)}_0 + Q^{(i)}_1 \lambda + \cdots Q^{(i)}_{n-i}\lambda^{n-i} \\
R^{(i)}(\lambda) &=& R^{(i)}_0 + R^{(i)}_1 \lambda + \cdots R^{(i)}_{i-1}\lambda^{i-1}
\end{eqnarray*}
Moreover, since $\psi_{n+1}=\mu$ we have to define
$$
Q^{(n+1)}(\lambda) = 0, \quad R^{(n+1)}(\lambda) = \prod_{j=1}^n (\lambda - \lambda_j)
$$
then eqs.(\ref{conditions}) become
$$
 \pmatrix{ 
1 & \lambda_1 & \cdots &\lambda_1^{i-1}& \mu_1 & \mu_1\lambda_1 & \cdots & \mu_1\lambda_1^{n-i-1}\cr
\vdots & \vdots &  &\vdots&\vdots & \vdots &  & \vdots\cr
1 & \lambda_j & \cdots &\lambda_j^{i-1}& \mu_j & \mu_j\lambda_j & \cdots & \mu_j\lambda_j^{n-i-1}\cr
\vdots & \vdots &  &\vdots&\vdots & \vdots &  & \vdots\cr
1 & \lambda_n & \cdots &\lambda_n^{i-1}& \mu_n & \mu_n\lambda_n & \cdots & \mu_n\lambda_n^{n-i-1}
}
\pmatrix{R^{(i)}_0 \cr \vdots  \cr R^{(i)}_{i-1}\cr Q^{(i)}_0 \cr \vdots \cr Q^{(i)}_{n-i-1} }
=  - Q^{(i)}_{n-i}  \pmatrix{\mu_1 \lambda_1^{n-i} \cr \vdots \cr \mu_j \lambda_j^{n-i} \cr \vdots \cr
\mu_n \lambda_n^{n-i} }
$$
or, with obvious notations $M^{(i)} X^{(i)} = - Q^{(i)}_{n-i} V^{(i)}$
and therefore $X^{(i)} = - Q^{(i)}_{n-i} M^{(i)-1}  V^{(i)}$.
In particular
$$
R^{(i)}_{k-1} = - Q^{(i)}_{n-i} {\Delta^{(i)}_k \over \Delta^{(i)} }
$$
 where 
$\Delta^{(i)} = \det M^{(i)}$ and $\Delta^{(i)}_k$ is the determinant of the matrix obtained from 
$M^{(i)}$ by replacing column $k$ by $V^{(i)}$.
Finally, one has to impose that the leading coefficients at $P_\pm$ are inverse to each other: 
$R_{i-1}^{(i)} = (Q^{(i)}_{n-i})^{-1}$. This gives
$$
(Q^{(i)}_{n-i})^{-2} = e^{q_i-q_{n+1}} = - {\Delta^{(i)}_i \over \Delta^{(i)} }
$$

To reconstruct the momenta, we follow \cite{BaBeTa03} again. Expand
$$
\psi_i = e^{q_i-q_{n+1}\over 2} \lambda^i ( 1 - \xi_i \lambda^{-1} + \cdots ), \quad {\rm near~} P^+
$$
then $ p_i = \xi_{i+1} - \xi_i $. We find at once
$$
\xi_i = -\sum_{j=1}^n \lambda_j - {R^{(i)}_{i-2} \over R^{(i)}_{i-1}}
$$
hence
$$
p_i = 
 {\Delta^{(i)}_{i-1} \over \Delta^{(i)}_{i}} - 
{\Delta^{(i+1)}_{i} \over \Delta^{(i+1)}_{i+1}}
$$
which we complement with the boundary terms
$$
p_1 = - {\Delta^{(2)}_{1} \over \Delta^{(2)}_{2}}, \quad p_n =  {\Delta^{(n)}_{n-1} \over \Delta^{(n)}_{n}} + \sum_{j=1}^n \lambda_j, \quad
p_{n+1} = - \sum_{j=1}^n p_j
$$

We now give more explicit formulae for the determinants entering the above expressions.
We call $[k]$ a subset of cardinality $k$ of  $(1,2,\cdots, n)$:
$$
[k]= (i_1,i_2,\cdots , i_k)
$$
We write $\sum_{[k]}$ for the sum over all such sets. Define
\begin{equation}
S_{[k]} = \prod_{i\in [k]}\prod_{j\neq [k]} {1\over (\lambda_{i} -\lambda_j)}
\label{defSk}
\end{equation}
and
$$
 \mu_{[k]} = \mu_{i_1}\mu_{i_2}\cdots \mu_{i_k}
 $$
 then, we have
\begin{eqnarray}
X^{(k)}&\equiv&  {\Delta^{(n-k)}\over \Delta^{(n)}} = \sum_{[k]}   S_{[k]} \mu_{[k]}
\label{defXr} \\
Y^{(k)}&\equiv&  {\Delta^{(n-k)}_{n-k-1}\over \Delta^{(n)}} =\sum_{[k]} S_{[k]} 
\left( \sum_{i\not\in [k]} \lambda_i \right)  \mu_{[k]} 
\label{defYr}
\end{eqnarray}

\bigskip
We have (note that $\Delta^{(i)}_i = (-1)^{n-i} \Delta^{(i-1)}$)

\begin{equation}
e^{q_i-q_{n+1}} ={X^{(n-i+1)}\over X^{(n-i)}},\quad p_i = {Y^{(n-i+1)}\over X^{(n-i+1)}}-{Y^{(n-i)}\over X^{(n-i)}}
\label{reconclass}
\end{equation}

It remains to check that the  Poisson bracket between $p_i,q_i$ is canonical. This easily follows from
\begin{proposition}
\begin{eqnarray*}
\{  X^{(k)}, X^{(l)} \} &=& 0 \\
\{ X^{(k)}, Y^{(l)} \} &=& (k-l) \theta(k-l)  X^{(k)} X^{(l)} \\
\{ Y^{(k)}, Y^{(l)} \} &=& (k-l) \Big(  \theta(k-l) Y^{(k)} X^{(l)}  + \theta(l-k)  X^{(k)} Y^{(l)}  \Big)
\end{eqnarray*}
where $\theta(k-l)=1\; {\rm if}\; k > l , 0 \;\rm{otherwise}$.
\end{proposition} 

Instead of proving these relations directly,
it is more convenient to use the quantity $Z^{(k)}$ defined by
\begin{equation}
Y^{(k)}= \left(\sum_{i=1}^n \lambda_i \right) X^{(k)} -Z^{(k)}, \quad  Z^{(k)} =\sum_{[k]} S_{[k]}  
\left( \sum_{i \in [k]} \lambda_i \right)\mu_{[k]} 
\label{defZr}
\end{equation}
Since $\{ \sum_{i=1}^n \lambda_i , X^{(k)}\} = - k X^{(k)}$, $ \{ \sum_{i=1}^n \lambda_i , Z^{(k)}\}  = - k Z^{(k)}$
we have to show that
\begin{proposition}
\begin{eqnarray*}
\{X^{(k)},X^{(l)}\}&=& 0\\
\{ X^{(k)}, Z^{(l)} \} &=&\Big( l \theta(k-l) + k \theta(l-k) \Big) X^{(k)} X^{(l)} \\
\{ Z^{(k)}, Z^{(l)} \} &=&  \Big( l \theta(k-l)+ k \theta(l-k) \Big)\Big( Z^{(k)} X^{(l)}  -   X^{(k)} Z^{(l)}  \Big)
\end{eqnarray*}
\end{proposition}
\proof
Take the semiclassical limit of the quantum formulae below.\square

Eqs.(\ref{reconclass}) and the above proposition provide a complete solution to the problem of expressing 
the original Toda variables $p_i,q_i$ in terms of the separated variables in the classical case. We now turn to the quantum theory.

\section{The Quantum Toda chain.}

In the quantum case, analysis on Riemann surfaces is not available. So, we try to quantize directly the 
relevant classical formulae. 

Quantum commutation relations are defined directly on the separated variables.
$$
[\lambda_k,\lambda_{k'}] =0,\quad \mu_k \lambda_{k'} = (\lambda_{k'} + i\hbar\delta_{kk'} ) \mu_k \equiv 
(t_k \lambda_{k'}) \mu_k,\quad
[\mu_k,\mu_{k'}] =0
$$
As shown in \cite{BaTa02}, the formulae (\ref{lagrangebis}) for the quantum Hamiltonians remain valid at the quantum level 
(with the $\mu_i$'s written on the right) and they are all commuting.

The new result of this paper concerns the variables $q_i,p_i$ of the Toda chain. We show that the classical formulae
eqs.(\ref{reconclass}) can also be straightforwardly quantized.

As a first step, we quantize  the operators $X^{(k)}$ and $Z^{(k)}$. We define them by the 
same formulae as in the classical case eqs.(\ref{defXr}, \ref{defZr}), but now it is important to write
the $\mu_i$'s to the right. We have
\begin{proposition}
\begin{eqnarray*}
[X^{(k)}, X^{(l)}] &=& 0 \\
\relax [X^{(k)}, Z^{(l)}] &=&   i\hbar (k \theta(l-k) + l\,\theta(k-l))   X^{(k)} X^{(l)}  \\
\relax [Z^{(k)}, Z^{(l)}] &=& i\hbar (k \theta(l-k) +  l\, \theta(k-l)  ) ( Z^{(k)}X^{(l)} - Z^{(l)}X^{(k)} )
\end{eqnarray*} 
\end{proposition}
\proof
We have
\begin{eqnarray*}
[X^{(k)}, X^{(l)}] &=& \sum_{[k],[l]} \Big( S_{[k]} t_{[k]} S_{[l]}- S_{[l]} t_{[l]} S_{[k]} \Big)\; \mu_{[k]}\mu_{[l]} \\
\relax [X^{(k)}, Z^{(l)}] &=& \sum_{[k],[l]} (S_{[k]} t_{[k]}(\lambda_{[l]} S_{[l]}) - \lambda_{[l]} S_{[l]}(t_{[l]}S_{[k]}) )\mu_{[k]}\mu_{[l]} \\
\relax [Z^{(k)}, Z^{(l)}] &=& \sum_{[k],[l]} (\lambda_{[k]}S_{[k]} (t_{[k]}\lambda_{[l]} S_{[l]}) - \lambda_{[l]} S_{[l]}(t_{[l]}\lambda_{[k]}S_{[k]}) )\mu_{[k]}\mu_{[l]}
\end{eqnarray*}
where we denoted
$$
\lambda_{[k]} = \sum_{i\in [k]} \lambda_i
$$
We set
\begin{equation}
[k]= [k']+[m'],\quad [l]=[l']+[m'],\quad [k']  \cap [l'] = \emptyset
\label{klm}
\end{equation}
We have
\begin{eqnarray*}
\sum_{[k],[l]}  ((\lambda_{[k]})^a S_{[k]} (t_{[k]}(\lambda_{[l]})^b S_{[l]}) - (\lambda_{[l]})^b S_{[l]}(t_{[l]}(\lambda_{[k]})^a S_{[k]})  &=&\sum_{[k],[l]} (-1)^{\#k' \#l'}\\
&& \hskip -5cm \times \prod_{i\in [k'+l'+m'] \atop j\not \in [k'+l'+m']} {1\over \lambda_i-\lambda_j} 
\prod_{i\in [m']\atop j \in [k'+l']} {1\over \lambda_i-\lambda_j}
 \prod_{i\in [m'] \atop j\not \in [k'+l'+m']} {1\over \lambda_i-\lambda_j+i\hbar} \\
&&\hskip -9cm \times
\prod_{i\in [k'] \atop j\in [l'] }{1\over \lambda_i-\lambda_j} 
\left((\lambda_{[k]})^a(\lambda_{[l]}+ i\hbar m')^b \prod_{i\in [k'] \atop j\in [l'] }{1\over \lambda_i-\lambda_j+ i\hbar}
- (\lambda_{[l]})^b(\lambda_{[k]}+ i\hbar m')^a \prod_{i\in [k'] \atop j\in [l'] }{1\over \lambda_i-\lambda_j- i\hbar}\right)
\end{eqnarray*}
The coefficient on the second line only depends on $[k'+l']$. Hence we can spilt the sum 
$$
\sum_{[k],[l]} = \sum_{[m'], [k'+l']}\; \sum_{[k'],[l']}
$$
and the last sum goes straight to the last line.

If $a=0,b=0$, that is in the calculation of $[X^{(k)}, X^{(l)}] $,  the last sum vanishes by Lemma \ref{lemma4}.

If $a=0,b=1$,  that is in the calculation of $[X^{(k)}, Z^{(l)}] $, we set
\begin{eqnarray*}
\lambda_{[l]}+ i\hbar m' &=&  \lambda_{[k'+l'+m']}- \lambda_{[k']} + i\hbar k - i\hbar k' \\
\lambda_{[l]} &=& \lambda_{[k'+l'+m']}- (\lambda_{[k']} - i\hbar k') - i\hbar k'
\end{eqnarray*}
By Lemma \ref{lemma4} applied with $a=0,1$, only the $i\hbar k$ term contributes in the last sum.
This term is exactly equal to
$$
[X^{(k)}, Z^{(l)}] = i\hbar k X^{(k)} X^{(l)},\quad k< l
$$
If $k>l$, we write this time
\begin{eqnarray*} 
\lambda_{[l]}+ i\hbar m' &=& ( \lambda_{[l']} - i\hbar l') + \lambda_{[m']} + i\hbar  l \\
 \lambda_{[l]} &=& \lambda_{[l']}+ \lambda_{[m']} 
\end{eqnarray*}
and this time only the $i\hbar l$ term contributes. Hence
$$
[X^{(k)}, Z^{(l)}] =i\hbar  l   X^{(k)} X^{(l)}
$$

If $a=1,b=1$,  that is in the calculation of $[Z^{(k)}, Z^{(l)}] $, we set (assuming $k < l$):
\begin{eqnarray*}
\lambda_{[k]}(\lambda_{[l]}+ m'i\hbar) &=& i\hbar k \lambda_{[k]} - (\lambda_{[k']})^2 ~~~~~~~+ (\lambda_{[k'+l']} -  i\hbar k') \lambda_{[k']}\\
&&\hskip 6cm + \lambda_{[m']}( \lambda_{[k'+l'+ m']}- i\hbar k') \\
\lambda_{[l]}(\lambda_{[k]}+ m'i\hbar) &=&i\hbar k \lambda_{[l]} - (\lambda_{[k']}-i\hbar k')^2 + (\lambda_{[k'+l']} - i\hbar  k') (\lambda_{[k']}-i\hbar k') \\
&&\hskip 6cm + \lambda_{[m']}( \lambda_{[k'+l'+ m']}- i\hbar k') 
\end{eqnarray*}
By Lemma \ref{lemma4} only the terms $i\hbar k \lambda_{[k]}$ and $i\hbar k \lambda_{[l]}$ contribute. Hence
$$
[Z^{(k)}, Z^{(l)}] =i\hbar   k  ( Z^{(k)}X^{(l)} - Z^{(l)}X^{(k)} ), \quad k< l
$$
\square

It is now simple to write the commutation relations with $Y^{(k)}$ defined in eq.(\ref{defYr}).
\begin{proposition}
\begin{eqnarray*}
[ X^{(k)},X^{(l)}] &=& 0 \\
 \relax [ X^{(k)},Y^{(l)}] &=&  i\hbar (k-l)  \theta (k-l) X^{(k)} X^{(l)}   \\
\relax [ Y^{(k)},Y^{(l)}] &=& i\hbar (k-l) \Big[ \theta(k-l) Y^{(k)} X^{(l)}+\theta(l-k) Y^{(l)} X^{(k)} \Big] 
\end{eqnarray*}
\end{proposition}

We define the quantum Toda variables as in the classical case:
\begin{proposition}
Let us define the quantum Toda operators as
\begin{equation}
e^{q_i-q_{n+1}} ={X^{(n-i+1)}\over X^{(n-i)}},\quad p_i = {Y^{(n-i+1)}\over X^{(n-i+1)}}-{Y^{(n-i)}\over X^{(n-i)}}
\label{reconquant}
\end{equation}
Then, we have
\begin{equation}
\left[e^{q_i} , e^{q_j} \right] =0,\quad 
\left[ e^{q_i} , p_j\right] =
 -i\hbar \delta_{ij}e^{q_i},\quad
 \left[ p_i, p_j \right] = 0
 \label{canquant} 
\end{equation}
\end{proposition}
\proof
Note that there is no ordering ambiguity in the expressions eq.(\ref{reconquant}).
The proof of the canonical commutation relations relies on
$$
\left[ {1\over X^{(k)}} , Y^{(l)} \right] = - i\hbar (k-l) \theta(k-l) {X^{(l)}\over X^{(k)}}
$$
which implies in turn
$$
 \left[ {Y^{(k)}\over X^{(k)}}, {Y^{(l)}\over X^{(l)}} \right] =0
$$
\square

Eqs.(\ref{reconquant}, \ref{canquant}) constitute the main result of this paper.
It is important to check the reality of our operators. The conjugation operation on the variables 
$\lambda_k,\mu_k$ was given to us  by Sklyanin:
\begin{eqnarray*}
\lambda_k^* &=& \lambda_k \\
\mu_k^* &=& \prod_{j\neq k} {\lambda_k - \lambda_j + i\hbar \over \lambda_k - \lambda_j }\;  \mu_k
\end{eqnarray*}
This conjugation rule is found by requiring that the Hamiltonians $H_j$ be self conjugate.
It is a simple exercise to check that the operators $X^{(k)}$, $Y^{(k)}$ are self conjugate, and therefore so are 
$p_i,q_i$.

\section{Conclusion.}

Eqs.(\ref{reconquant}) opens up the possibility to compute the matrix elements of these operators
between the eigenstates of the Hamiltonians $H_i$, see \cite{Smi98,KhLe99}. The existing techniques 
should be sufficient to handle operators polynomials in $\mu_i$, such as
$$
e^{q_n-q_{n+1}}e^{q_{n-1}-q_{n+1}}\cdots e^{q_{n-k+1}-q_{n+1}} = X^{(k)}
$$
For the operators $p_i, e^{q_i}$ themselves however, we will have to learn how to treat ratios of such
polynomial operators. This is an important issue since for non hyperelliptic curves this situation seems 
unavoidable \cite{SmiZe02}.

\section{Appendix.}
We prove some combinatorial identities which are used in the computation of the commutators of the 
operators $X^{(k)}$, $Z^{(k)}$.
\begin{lemma}
$$
\sum_{i=1}^n  \prod_{j\neq i} {1\over (\lambda_i-\lambda_j) } 
\left( \lambda_i^a \prod_{j\neq i} {1\over (\lambda_i  -\lambda_j+ i\hbar) }
- (\lambda_i-i\hbar)^a\prod_{j\neq i} {1\over (\lambda_i -\lambda_j - i\hbar) } \right)=0
$$
for $a \leq 2n$.
\label{lemma3}
\end{lemma}
\proof
Consider
$$
Q(z) = z^a\prod_i {1\over z-\lambda_i} \prod_i{1\over z - \lambda_i  + i\hbar }
$$
The residue at the pole $z = \lambda_i$ reads
$$
{1\over i\hbar}\lambda_i^a \prod_{i\neq j} {1\over \lambda_i-\lambda_j} \prod_{i\neq j}{1\over \lambda_i - \lambda_j  + i\hbar }
$$
while the residue at $z = \lambda_i -i\hbar$ reads
$$
-{1\over i\hbar}(\lambda_i -i\hbar)^a \prod_{i\neq j} {1\over \lambda_i-\lambda_j-i\hbar} \prod_{i\neq j}{1\over \lambda_i - \lambda_j }
$$
Hence our expression is the sum of the residues of $Q(z)$ at finite distance, which vanishes if $a\leq 2n$.
\square

\begin{lemma}
Suppose $k < n/2$. Then
$$
\sum_{[k]}  \prod_{i \in [k] \atop j \not\in [k]} {1\over (\lambda_i-\lambda_j) } 
\left((\lambda_{[k]})^a \prod_{i \in [k] \atop j \not\in [k]} {1\over (\lambda_i  -\lambda_j+ i\hbar) }
- (\lambda_{[k]}-  i\hbar k)^a\prod_{i \in [k] \atop j \not\in [k]} {1\over (\lambda_i -\lambda_j - i\hbar) } \right)=0
$$
for $0 \leq a \leq 2(n-2k+1)$. By symmetry, if $k>n/2$, we have
$$
\sum_{[k]}  \prod_{i \in [k] \atop j \not\in [k]} {1\over (\lambda_i-\lambda_j) } 
\left((\lambda_{[n-k]}-  i\hbar (n- k))^a \prod_{i \in [k] \atop j \not\in [k]} {1\over (\lambda_i  -\lambda_j+ i\hbar) }
- (\lambda_{[n-k]})^a\prod_{i \in [k] \atop j \not\in [k]} {1\over (\lambda_i -\lambda_j - i\hbar) } \right)=0
$$
for $0 \leq a \leq 2(2k-n+1)$
\label{lemma4}
\end{lemma}
\proof
Consider this expression as a function of $\lambda_1$. It tends to zero at $\infty$, and
it has poles at the other $\lambda_j, \lambda_j \pm i\hbar$. Consider $\lambda_2$. We have two contributions
corresponding to $\lambda_1 \in [k], \lambda_2 \not\in [k]$ and $\lambda_1 \not\in [k], \lambda_2 \in [k]$.
We denote by $[n']$ the subset of $[n]$ where $\lambda_1$ and $\lambda_2$ have been removed, by $[k']$
a subset of $[n']$ of cardinality $k-1$, and by $[l']$ the complementary subset in $[n']$.
The two contributions can be written respectively:
\begin{eqnarray*}
A &=& {1\over \lambda_1 - \lambda_2} P_{[l']}(\lambda_1)P_{[k']}(\lambda_2) {\prod}_0'
\left( {(\lambda_1+\lambda_{[k']})^a\over \lambda_1 - \lambda_2 + i\hbar}  P_{[l']}(\lambda_1+i\hbar)P_{[k']}(\lambda_2-i\hbar)  {\prod}_+'
\right. \\
&&\hskip 4cm \left. -{(\lambda_1+\lambda_{[k']}- k i\hbar)^a\over \lambda_1 - \lambda_2 - i\hbar}  P_{[l']}(\lambda_1-i\hbar)P_{[k']}(\lambda_2+i\hbar)  {\prod}_-' \right)
\end{eqnarray*}
\begin{eqnarray*}
B &=& {1\over \lambda_2 - \lambda_1} P_{[l']}(\lambda_2)P_{[k']}(\lambda_1) {\prod}_0'
\left( {(\lambda_2+\lambda_{[k']})^a\over \lambda_2 - \lambda_1 + i\hbar}  P_{[l']}(\lambda_2+i\hbar)P_{[k']}(\lambda_1-i\hbar)  {\prod}_+'
\right. \\
&&\hskip 4cm \left. -{(\lambda_2+\lambda_{[k']}- k i\hbar)^a\over \lambda_2 - \lambda_1 - i\hbar}  P_{[l']}(\lambda_2-i\hbar)P_{[k']}(\lambda_1+i\hbar)  {\prod}_-' \right)
\end{eqnarray*}
where 
$$
P_{[l']}(\lambda) = \prod_{j\in [l']} {1\over \lambda - \lambda_j}
$$
and 
$$
{\prod}_\sigma'= \prod_{i \in [k'] \atop j \in [l']} {1\over (\lambda_i-\lambda_j + \sigma i\hbar) }, \quad \sigma = 0,\pm 
$$

Consider the pole at $\lambda_1 = \lambda_2$. Set $\lambda_1=\lambda_2 + \epsilon$.
\begin{eqnarray*}
A &=& {1\over \epsilon} P_{[l']}(\lambda_2) P_{[k']}(\lambda_2)  {\prod}_0'
\left( {1\over  i\hbar}(\lambda_2+\lambda_{[k']})^a  P_{[l']}(\lambda_2+ i\hbar)  P_{[k']}(\lambda_2-i\hbar){\prod}_+'
\right. \\
&&\hskip 3cm \left. +{1\over  i\hbar}(\lambda_2+\lambda_{[k']}- k i\hbar)^a  P_{[l']}(\lambda_2-i\hbar ) P_{[k']}(\lambda_2+i\hbar) {\prod}_-' \right)
\end{eqnarray*}
\begin{eqnarray*}
B &=& -{1\over \epsilon}  P_{[l']}(\lambda_2) P_{[k']}(\lambda_2)  {\prod}_0'
\left( {1\over  i\hbar} (\lambda_2+\lambda_{[k']})^a P_{[l']}(\lambda_2+ i\hbar)  P_{[k']}(\lambda_2-i\hbar){\prod}_+'
\right. \\
&&\hskip 3cm \left.+ {1\over  i\hbar}(\lambda_2+\lambda_{[k']}- k i\hbar)^a  P_{[l']}(\lambda_2-i\hbar ) P_{[k']}(\lambda_2+i\hbar) {\prod}_-' \right)
\end{eqnarray*}
so that $A+B$ is regular.\\

Consider the pole at $\lambda_1 = \lambda_2 - i\hbar$. Set $\lambda_1 = \lambda_2 - i\hbar + \epsilon$
\begin{eqnarray*}
A &=& -{1\over i\hbar\epsilon}P_{[n']}(\lambda_2-i\hbar) P_{[n']}(\lambda_2) 
  (\lambda_2-i\hbar+\lambda_{[k']} )^a {\prod}_0' {\prod}_+' \\
 B &=& {1\over i\hbar\epsilon} P_{[n']}(\lambda_2) P_{[n']}(\lambda_2-i\hbar) 
 (\lambda_2-i\hbar+\lambda_{[k']}- (k-1) i\hbar)^a{\prod}_0'  {\prod}_-' 
\end{eqnarray*}
so that $A+B$ is proportional  to 
$$
\sum_{[k']} {\prod}_0'\left((\lambda_{[k']})^{a'}{\prod}_+'-
(\lambda_{[k']}- (k-1)i\hbar)^{a'}{\prod}_-'\right)
$$
which is our identity at a lower level.\\

Consider the pole at $\lambda_1 = \lambda_2 + i\hbar$. Set $\lambda_1 = \lambda_2 + i\hbar + \epsilon$.
\begin{eqnarray*}
A &=& -{1\over i\hbar \epsilon}P_{[n']}(\lambda_2+i\hbar) P_{[n']}(\lambda_2)  
(\lambda_2+\lambda_{[k']}-(k-1)i\hbar )^a {\prod}_0' {\prod}_-' \\
B &=& {1\over i\hbar\epsilon} P_{[n']}(\lambda_2) P_{[n']}(\lambda_2+i\hbar)  
(\lambda_2+\lambda_{[k']} )^a {\prod}_0' {\prod}_+' 
\end{eqnarray*}
so that $A+B$ is proportional  to 
$$
\sum_{[k']}{\prod}_0'\left((\lambda_{[k']})^{a'}{\prod}_+'-
(\lambda_{[k']}- (k-1)i\hbar)^{a'}{\prod}_-'\right)
$$
 which is our identity at a lower level. So, if the identity holds at lower levels, our expression, as a rational function of
 $\lambda_1$, is regular everywhere and tends to zero at $\infty$, hence identically vanishes. Since the identity  is true for 
 $k=1$ by Lemma \ref{lemma3}, it is true as stated.
\square

{\bf Acknowledgements.} I thank  F. Smirnov and D. Talalaev for discussions.

 \end{document}